\documentclass{article}

\usepackage{arxiv}

\usepackage[utf8]{inputenc} 
\usepackage[T1]{fontenc}    
\usepackage{hyperref}       
\usepackage{url}            
\usepackage{booktabs}       
\usepackage{amsfonts}       
\usepackage{nicefrac}       
\usepackage{microtype}      
\usepackage[english]{babel}
\usepackage{fancyhdr}

\usepackage{svg}
\usepackage{subfigure}
\usepackage{multirow} 
\usepackage{booktabs}
\usepackage{amsmath}
\usepackage{amssymb}
\usepackage{academicons}
\usepackage{xcolor}
\usepackage{orcidlink}

\newcommand{\citep}{\cite}
\newcommand{\citet}{\cite}

\title{Metric Learning for Session-based Recommendations}

\author{
Bartłomiej Twardowski\textsuperscript{1,2} \orcidlink{0000-0003-2117-8679} \\
\textsuperscript{1}Computer Vision Center,\\ 
Universitat Autónoma de Barcelona \\
\textsuperscript{2}Warsaw University of Technology, \\ 
Institute of Computer Science\\
\texttt{bartlomiej.twardowski@pw.edu.pl}
\And
Paweł Zawistowski \orcidlink{0000-0002-0273-7060} \\
Warsaw University of Technology, \\ 
Institute of Computer Science\\
\texttt{pawel.zawistowski@pw.edu.pl}
\And
Szymon Zaborowski\\
Sales Intelligence
}

\begin{document}
\maketitle

\begin{abstract}
Session-based recommenders, used for making predictions out of users' uninterrupted sequences of actions, are attractive for many applications. Here, for this task we propose using metric learning, where a common embedding space for sessions and items is created, and distance measures dissimilarity between the provided sequence of users' events and the next action. We discuss and compare metric learning approaches to commonly used learning-to-rank methods, where some synergies exist. We propose a simple architecture for problem analysis and demonstrate that neither extensively big nor deep architectures are necessary in order to outperform existing methods. The experimental results against strong baselines on four datasets are provided with an ablation study. 
\end{abstract}


\section{Introduction}

We consider the session-based recommendation problem, which is set up as follows: a user interacts with a given system (e.g., an e-commerce website) and produces a sequence of events (each described by a set of attributes). Such a continuous sequence is called a session, thus we denote $s_{k} = e_{k,1},e_{k,2},\ldots,e_{k,t}$ as the $k$-th session in our dataset, where $e_{k,j}$ is the $j$-th event in that session. The events are usually interactions with items (e.g., products) within the system's domain. In comparison to other recommendation scenarios, in the case of session-based recommendations --- information about the user across sessions is not available (in contrast to session-aware recommendations). Also, the browsing sessions originate from a single site (which is different from task-based recommendations).

The sequential nature of session-based recommendations means that it shares some similarities with tasks found within natural language processing (NLP), where sequences of characters, words, sentences, or paragraphs are analyzed. This connection leads to a situation where many methods that are successful in NLP are later applied to the field of recommendations. One such example is connected with recurrent neural networks (RNNs), which have led to a variety of approaches applied to recommender systems \cite{Hidasi,twardowski2016modelling,Smirnova2017rnn}.  Another, one is connected with the transformer model \cite{devlin2018bert} applied to model users' behavior \cite{sun2019bert4rec}.

Despite the apparent steady progress connected with neural methods, there are some indications that properly applied classical methods may very well beat these approaches \cite{Ludewig2019}.
Therefore in this paper, we propose combining the classical KNN algorithm with a neural embedding function based on an efficient neighborhood selection of top-n recommendations. The method learns embeddings of sessions and items in the same metric space, where a given distance function measures dissimilarity between the user's current session, and next items. For this task, a metric learning loss function and data sampling are used for training the model. During the evaluation, the nearest neighbors are found for the embedded session. This makes the method attractive for real-life applications, as existing tools and methods for neighborhood selection can be used. The main contributions of this paper are as follows:

\begin{itemize}
    \item we verify selected metric learning tools for session-based recommendations,
    \item we present a comparison of the metric learning approach and learning to rank, where some potential future directions for recommender systems can be explored based on the latest progress in deep metric learning,
    \item we introduce a generic model for recommendations, which allows the impact of different architectures of session and item encodings on the final performance to be evaluated--- which we do in the provided ablation studies,
    \item we evaluate our approach using known protocols from previous session-based recommendation works against strong baselines over four datasets; for the sake of reproducibility and future research\footnote{https://github.com/btwardow/dml4rec}.
\end{itemize}

\section{Related Works}

\subsection{Session-based Recommendations}
Time and sequence models in context-aware recommender systems were used before the deep learning methods emerged. Many of these approaches can be applied to session-based recommendation problems with some additional effort to represent time, e.g., modeling it as a categorical contextual variable~\cite{hidasi2016general,Rendle2010FM} or explicit bias while making predictions~\cite{Koren2009td}. The sequential nature of the problem can also be simplified and used with other well-known methods, i.e., Markov chains~\cite{Rendle2010FMC}, or applying KNNs combined with calculating the session items sets' similarities \cite{Jannach2017}. 

The Gru4Rec method \cite{Hidasi} has been an important milestone in applying RNNs to session-based recommendation tasks. The authors focused on sessions solely represented by interactions with items and proposed a few contributions: using GRU cells for session representation, negative exemplars mining within mini-batch, and a new \texttt{TOP1} loss function. In the followup work \cite{hidasi2018recurrent} authors proposed further improvements to loss functions. Inspired by the successful application of convolutional neural networks (CNNs) for textual data~\cite{kim2014convolutional}, new methods were proposed. One example is the Caser approach \cite{Tang2018caser}, which uses a CNN-based architecture with max pooling layers for top-n recommendations for sessions. Another, proposed in \cite{Yuan2019}, utilises dilated 1D convolutions similar to WaveNet \cite{Oord2016}. The embedding techniques known from NLP, e.g. skip-gram and CBOW, were also extensively investigated for recommender systems. Methods such as item2vec and prod2vec were proposed for embedding-based approaches. However recently conducted experiments with similar approaches, were unsuccessful in obtaining better results than simple neighbourhood methods for session-based recommendations \cite{Ludewig2018eval}.  

\subsection{Metric Learning} 
Metric learning has a long history in information retrieval. Among the early works, the SVM algorithm was used to learn from relative comparisons in \cite{schultz2004learning}. Such an approach directly relates to Mahalanobis distance learning, which was pursued in \cite{mcfee2010metric} and \cite{lim2014efficient}. Even though new and more efficient architectures emerge constantly, the choice of loss functions and training methods still plays a significant role in metric learning. In \cite{hadsell2006dimensionality}, the authors proposed the use of contrastive loss, which minimizes the distance between similar pairs while ensuring the separation of non-similar objects by a given margin. For some applications, it was found hard to train, and in \cite{Hoffer2015}, the authors proposed improvement by using an additional data point---\emph{anchor}. All three data points make an input to the triplet loss function, where the objective is to keep the negative examples further away from the anchor than the positive ones with a given margin. Recently more advanced loss functions were proposed: using angular calculation in triplets\cite{wang2017angular}, signal-to-noise ratio\cite{yuan2019signal}, and multi-similarity loss\cite{wang2019multi}. Still, contrastive and triple losses in many applications have proven to be a strong baseline when trained correctly\cite{hermans2017defense}.
Nevertheless, the high computational complexity of data preparation (i.e. creating point tuples for training) for contrastive and triplet approaches cannot be solved by changing only the loss function. These problems are addressed by different dataset sampling strategies and efficient mining techniques. One notable group here is online approaches, which try to explore relations in a given mini-batch while training, e.g., hard mining\cite{hermans2017defense}, n-pairs\cite{sohn2016npairs}, the lifted structure method\cite{oh2016lifed}, and weighting by distance\cite{wu2017sampling}. Many combinations of sampling and mining techniques, along with the loss functions, can be created, which makes a fair comparison hard \cite{fehervari2019unbiased,musgrave2020realitycheck,kaya2019dmlsurvey}.

\section{Metric Learning vs Ranking Learning for Session-based Recommendations}

An ordered output of the session-based recommender in the form of a sorted list for a given input $s_k$ is the ranking $r_k$. In learning-to-rank, as well as recommender systems, the main difficulty is the direct optimization of the output's quality measures (e.g., recall, mean average precision, or mean reciprocal rank). The task is hard for many (gradient-based) methods due to the non-smoothness of the optimized function\cite{Burges2007}. This problem can be resolved either by minimizing a convex upper bound of the loss function, e.g., SVM-MAP\cite{Yue2007}, or by optimizing a smoothed version of an evaluation measure, e.g.,~SoftRank\cite{Taylor2008}. Many approaches exist, which depend on the model of ranking: pointwise (e.g.~SLIM\cite{Ning2011}, MF\cite{Koren2008}, FM\cite{Rendle2010FM}), pairwise (BPR\cite{Rendle2009}, pLPA\cite{Liu2009}, GRU4Rec\cite{Hidasi}), or list-wise (e.g. CLIMF/xCLIMF \cite{Chen2006,Shi2013}, GAPfm\cite{GAPfm}, TFMAP\cite{Shi2012}). However, not all are applicable to session-based recommendations. Pairwise approaches for ranking top-N items are the most commonly used, along with neural network approaches. In the GRU4Rec method, two pairwise loss functions for training were used --- \emph{Bayesian Personalized Ranking} (BPR)\cite{Rendle2009} and TOP-1:
\begin{align}
l_\text{BPR} (s_k, i_p, i_n) &= -\ln \bigl( \sigma ( \hat{y}_{s_k,i_p} - \hat{y}_{s_k,i_n} ) \bigr)\\
l_\text{TOP1} (s_k, i_p, i_n) &= \sigma ( \hat{y}_{s_k,i_n} - \hat{y}_{s_k,i_p} ) + \sigma (\hat{y}_{s_k,i_n}^{2})
\end{align}
where $s_k$ denotes the session for which $i_p$ is a positive example of the next item, and $i_n$ is a negative one. The $\hat{y}_{s_k,i}$ is a score value predicted by the model for the session $s_k$ and the item $i$. The score value allows items to be compared and the ordered list $r_k$ to be produced, where i.e. $i_p >_{r_k} i_n$, and $>_{r_k} \subset I^2$ denotes total order\cite{Rendle2009}.

In metric learning, the main goal is to align distance with dissimilarity of objects. In \cite{hadsell2006dimensionality}, the contrastive loss function for two vectors $\mathbf{x}_i, \mathbf{x}_j \in \mathbb{R}^{\mathtt d}$ is given as:
 \begin{equation}
    l_\text{Cont} (\mathbf{x}_i, \mathbf{x}_j) = y d(\mathbf{x}_i, \mathbf{x}_j) + (1-y) \max\bigl(0, d(\mathbf{x}_a, \mathbf{x}_n) - m\bigr)
\end{equation}
where $y$ is an indicator variable, 1 if both vectors are from the same class, 0 otherwise, $m \in R_{+}$ is the margin, and $d(\mathbf{x_i}, \mathbf{x_j})$ is a distance function, e.g., Euclidian 
or cosine. This loss function \emph{pulls} similar items ($y=1$) and \emph{pushes} dissimilar ones. A direct extension –- the triplet loss\cite{Hoffer2015} -- is defined as follows: 
\begin{equation}
    l_\text{Triplet} (\mathbf{x}_a, \mathbf{x}_p, \mathbf{x}_n) = \max\bigl(0,\   d(\mathbf{x}_a, \mathbf{x}_p) - d(\mathbf{x}_a, \mathbf{x}_n) + m \bigr)    
\end{equation}
where $\mathbf{x}_p$ and $\mathbf{x}_n$ are respectively positive and negative items for a given anchor $\mathbf{x}_a$ and $m \in R_{+}$ is the margin. 

Both contrastive and triplet losses can be used to optimize the goal of the total ordering of objects \cite{liu2009learning,Rendle2009} as induced by the learned metric. If $d(\mathbf{x_i}, \mathbf{x_j}) = 0$ does not imply $\mathbf{x_i} = \mathbf{x_j}$, $d$ is then a pseudo metric \cite{schultz2004learning}, and total order cannot be induced. If we assume that two functions $\varphi(s_a)=x_a$ and item $\omega(i_{k})=x_{k}$ are given to embed the session and item to the same $\mathbb{R}^{\mathtt{d}}$ space, where scoring is done by cosine similarity $\hat{y}_{s,i} = 1 - d(\varphi(s), \omega(i))$, then previously defined ranking losses and metric can be presented as: 
\begin{align}
l_\text{BPR} (s_k, i_p, i_n) &= -\ln ( \sigma ( d_{kn} - d_{kp}))\ , \\
l_\text{TOP1} (s_k, i_p, i_n) &= \sigma (d_{kp} - d_{kn}) + \sigma((1 - d_{kn})^{2})\ , \\
l_\text{Triplet} (s_k, i_p, i_n) &= \max(0, d_{kp} - d_{kn} + m )   
\end{align}
where $d_{kj} = d(\varphi(s_k), \omega(i_j))$.
A direct connection can be seen: that minimizing each of the loss functions will try to keep $i_p$ closer to $s_k$ than $i_n$. In all cases, for session-based recommendations, positive items are known, while the negatives are sampled from the rest of the items (e.g., uniformly or by a given heuristic). In both \texttt{BPR} and \texttt{TOP1}, a sigmoid $\sigma(x)$ function is used for optimizing AUC in place of a non-differentiable Heaviside function directly, as explained in \cite{Rendle2009}. In \texttt{TOP1}, the authors added a regularization term for negative predictions, which further constrains the embedding space by keeping negatives close to zero. Metric learning losses use a rectifier nonlinearity ($\max(0,x)$) to prevent from moving data points that are already in order. When considering partial derivative w.r.t distances between our anchor session $s_k$ and positive and negative items, they contribute equally, as was discussed in \cite{wang2019multi}. If in a single calculation, more relations are explored (usually inside the same mini-batch), techniques like lifted structures \cite{oh2016lifed} are used. However, the relations are made between known classes of examples. In learning to rank, each instance inside a selected set can be ordered, which can be used i.e., to estimate overall ranking, like in Weighted Approximated-Ranking Pairwise (WARP)\cite{weston2011wsabie}. All losses have one more important thing in common: they do not take into account the relationship between positive and negative items (without the anchor). This is a subject of further improvements in metric learning methods like \cite{wang2017angular,wang2019multi}. In our solution, we propose using a simple weighting for ranking to address this shortcoming.

\section{Proposed method}

We propose a method for session-based recommendations using deep metric learning, where the main input is the sequence of user's actions (i.e. the session) $s_k = \{ e_{k,1},e_{k,2},\ldots,e_{k,t} \} \in S$, and items $i \in I$. At the high-level the network's architecture can be described as $\hat y_{s_k,i} = d(\varphi(s_k), \omega(i))$, where $\varphi$ and $\omega$ denote the session and item encoders respectively, and $\hat y_{s_k,i}$ denotes how score for recommending item $i$ in the context of session $s_k$. We decided on a simple and modular approach in order to investigate the impact of each module on the final outcome --- focusing mainly on the session encoder and different metric loss functions. The only constraint of the model towards the used network is the used dimensionality of $\varphi(s_k),\ \omega(i) \in \mathbb{R}^{\mathtt d}$ for learning a common metric space. The outputs of networks are normalized and cosine distance functions $d(\varphi(s_k), \omega(i))$ are used in final scoring $\hat{y}_{s_k, i}$ calculation.

\subsection{Metric Loss for Ranking}

\subsection{Triplet Loss} The overall triplet loss function is calculated over the prepared training dataset. Assuming that session $s_k$ has $L$ positive items, the final triplet loss function for balanced positive-negative sampling is as follows: 
\begin{equation}
L = \frac{1}{|S|} \sum_{s_k \in S} \sum_{j=0}^{L} w_{j} \max\left(0, d \left(\varphi(s_k), \omega(i_p)\right) - d\left(\varphi(s_k), \omega(i_n)\right) + m \right)
\end{equation}
where weight $w_j$ is weight used for particular position. In experiments, we used $\sqrt{1 /(1 + j)}$ for weighting, which is expected to change the magnitude of the calculated gradient based on the ranking position. To incorporate the relation between positives and negatives items we used a \emph{swaping} technique for a triplet loss, where anchor is exchanged with positive and the final distance to a negative point is taken as a minimum $d'_{kn} = min\left( d_{kn}, d_{pn} \right)$.

\subsection{Neighborhood Component Analysis with Smoothing (NCAS) Loss} 
Based on the NCA loss \citep{goldberger2005nca,movshovitz2017proxy-nofuss} used commonly in deep metric learning we introduce a version prepared for ranking session-based recommendations as follows:
\begin{align}
    p(i_j | s_k) &= \frac{exp\left( - d\left(\varphi(s_k), \omega(i_j)\right) \right)} {\sum_{i_j \in Z} exp\left( - d(\varphi(s_k), \omega(i_j)) \right)} \\
    L_{NCAS} &=  \frac{1}{|S|} \sum_{s_k \in S} KLD \left( p(i | s_k) || p'(i) \right)
\end{align}
where predictions of true labels inside $N$-sized mini-batches are smoothed with: $p'(i) = (1 - \epsilon)p(i) + \epsilon / N$ and $Z$ is a sampled set containing positive and negative examples for each session $s_k$. The main goal of using this loss function was to compare the triplet loss to other functions that can be applied for our setting in order to get more insight of its applicability and results.

\subsection{Session Encoder Networks}
\label{sec:encoders}

We use several neural network architectures for the session encoder module. Each one of these networks takes as an input a sequence of session events, which are clicked items in all used datasets, and encodes it to a vector of embedding size $d$. Used network architectures as session encoders go as follows:
\vspace{-0.5em}
\begin{itemize}
    \item Pooling - this architecture embeds the sequence of clicked items to a vector of size $d$ by pooling the maximum or average value in each dimension.
    Inspired by how pre-trained embeddings (e.g. word2vec) are used in NLP downstream tasks. However, all relations in a sequence are lost.
    \item CNN based approaches including TextCNN\cite{kim2014convolutional}, TagSpace\cite{weston-etal-2014-tagspace}, Caser\cite{Tang2018caser}. 
    \item RNN-based approaches --- these use one of the chosen recurrent networks (GRU, LSTM, RNN) to encode the sequence followed by multiple fully connected layers to generate recommendation scores for individual items.
\end{itemize}

\subsection{Positive and negative sampling}
\label{sec:sampling}

Training data is prepared from all available users' sessions $S$. We want to predict the user's next action for a given session $s_k$. Thus, training data preparation tries to enforce this for the model. Each session is split randomly --- the first part is used as an input for the network $s_k$, and the following actions with items are used as positive examples for that session $i_{p,1},\ldots,i_{p,l}$. For each $l$ positive, the same number of negatives are sampled randomly. We investigated a few different strategies in case the session after random split has not enough positive examples. One of the successful approaches we used is to prepare more positives before training using KNN method. Sampling is done at a beginning of each training epoch. However, the improved MRR score is counterbalanced by lower items' coverage.

In other works, the negative sampling is done randomly from all non-positive items, e.g., \cite{Rendle2010FM}. From the optimization perspective \citet{Hidasi} took a different approach and sample negative examples from the same mini-batch given to the network. What relates to online samples mining used in deep metric learning techniques, but here without enforcing a margin of error like in hard negative mining \cite{hermans2017defense}.

\section{Experiments}

To conduct our experiments, we have followed the procedure utilized by \citet{Ludewig2019}, and used five splits for \texttt{RR} and one 64'th of \texttt{RSC15} data. For each dataset, we have split the events into individual user sessions and removed the ones that contained only a single event. Furthermore, in our experiments, we have included only items that occurred at least five times in the data. A train-test split was prepared by taking the last $10\%$ of sessions. We further evaluated our models by using common information retrieval and ranking evaluation metrics: mean average precision, mean reciprocal rank, recall, precision, and hit ratio. All metrics were computed on a list of top 20 recommendations. Following \citet{Ludewig2019} and \citet{Hidasi}, in case of MRR@20 and HR@20 only the next item was used as the ground truth. This \emph{no look-ahead} evaluation can be considered as a more adequate, when after each of a user's action a session state is updated and predictions for the next user's step is given.

\subsection{Datasets and baselines}

To conduct the experiments, we used four datasets from the e-commerce domain, which are summarized in Table~\ref{tab:datasets}.  Two of these (\texttt{RR/5} and \texttt{RSC15/64}) are standard benchmarks for session-based recommender systems, while the remaining ones are smaller, real-world proprietary datasets with data gathered in the early 2020. The difference between \texttt{SI-T} and \texttt{SI-T} is the category of products for which data were collected. In all datasets users' events are represented only by interactions with a products (i.e.,~view, click), thus $e_{k,l} \sim I$.

\begin{figure}[tp]
    \centering
    \begin{tabular}{cc}
    \subfigure{\includegraphics[scale=1]
    {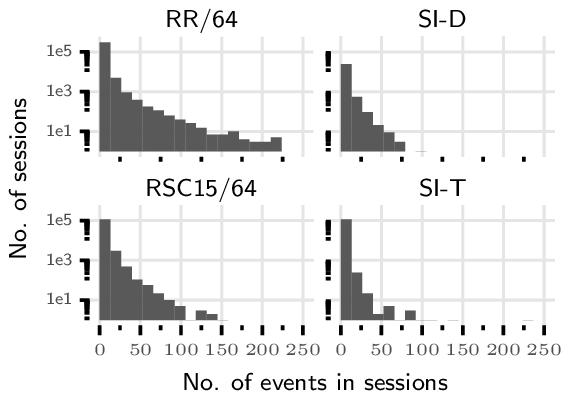}} &
    \subfigure{\includegraphics[scale=1]{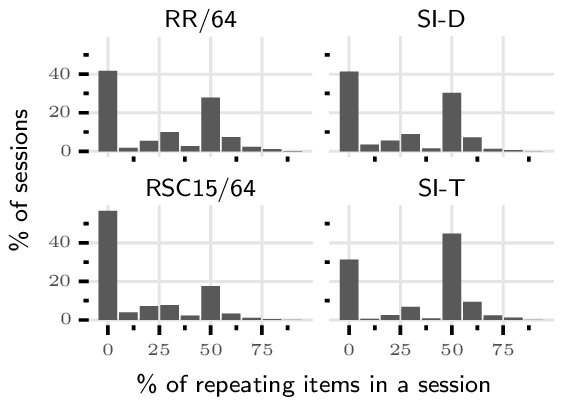}}\\
    \end{tabular}
    \caption{Session length distribution (left) and repeating items (right) for each dataset.}
    \label{fig:preprocessed_session_lengths}
\end{figure}

\begin{table}[t]
    \centering
    \small
    \begin{tabular}{r|l|rr|rr|rr}
         Dataset         &  Source 
         & \multicolumn{2}{c|}{Items}&  \multicolumn{2}{c|}{Sessions} & \multicolumn{2}{c}{Events}\\
         \hline
         \texttt{RR/5}     &  Retail Rocket
         & 32K & (117K) 
         & 64K & (380K)
         & 242K & (606K) \\
         
         \texttt{RSC15/64}  & 2015 RecSys Challange
         & 17K & (34K) 
         & 118K & (1.7M)   
         & 495K & (6.6M)\\
         
         \texttt{SI-T}    & Proprietary e-commerce data \#1
         & 2K & (2K) 
         & 114K  & (119K)
         & 305K &  (315K)\\
         
         \texttt{SI-D}    & Proprietary e-commerce data \#2
         & 3K  & (3K)
         & 25K & (34K)
         & 94K & (106K) \\
    \end{tabular}
    \caption{Experimental dataset stats --- \textbf{(}before\textbf{)} and after preprocessing.}
    \label{tab:datasets}
\end{table}
Fig.~\ref{fig:preprocessed_session_lengths} (left) presents a histograms of session lengths for the preprocessed datasets, which shows that short sessions seem to dominate in all datasets. This might be more challenging for methods that focus on the sequential nature of the users' data. Furthermore, when analyzing the percentage of recurring items among the sessions presented on Fig.~\ref{fig:preprocessed_session_lengths} (right),  it may be noticed that the session frequently contain multiple interactions with the same products. The data suggests that users seem to revisit already seen items quite often. However, this also poses an interesting question from the perspective of recommender systems: should such a system suggest items that a user has already seen in the given session or only new ones? The answer will depend on the specific use case and whether the system should provide a more explorative or exploitative user experience.

We compared our Session-based Metric Learning (\texttt{SML}) method against six baseline algorithms. Starting from the simplest ones, \texttt{POP} denotes a simple pop\-ularity-based algorithm, which simply recommends the top-n most popular items. \texttt{SPOP} recommend items already seen in the session ordered by number of occurrences and fills the rest with popular ones. This recommender performs well when predictions are expected to be repetitions. The \texttt{KNN} algorithm was the basis of the next two baseline methods: \texttt{SKNN} and \texttt{VSKNN}. The \texttt{SKNN} approach for a given session recommends the top-n most frequent items among the \texttt{K}-most similar sessions from the training data, for which a cosine distance is used. The \texttt{VSKNN}\cite{Ludewig2019} approach works similarly, however it puts more weight on more recent events in a given session. The last two methods include a Markov first-order recommender reported as \texttt{MARKOV-1} and \texttt{GRU4Rec+}\cite{hidasi2018recurrent}.

\subsection{Implementation details}

All the variations of the proposed model were implemented using the PyTorch\cite{pytorch} library and trained in an end-to-end fashion with Adam optimizer using $lr=0.001$, for max 150 epochs (early stop after lowering $lr$ three times when improvement on $5\%$ validation data is lower than $0.5\%$) with batch size of 32, and 8 positive/negative samples per session. Max session length was 15 for \texttt{RR/5} and \texttt{RSC15/64}, and 8 for SI --- this plays an important role for CNNs where all sessions are padded to exactly the same size. For item embedding simple feed-forward network with $tanh$ activation is used. The embedding dimension is set to 400 for all the methods. The margin value $m$ for triplet loss is set to $0.3$. Smoothing parameter for NCAS is set to $\epsilon = 0.3$. For the \texttt{RNN} encoder a GRU cells are used with, 400 dimensions. For the \texttt{TextCNN} convolution filters of sizes $1,3,5$ were used.

\subsection{Performance comparison}

\subsection{Evaluation} In Table~\ref{tab:experimental_results} we present the results obtained during the experiments conducted with the proposed method and compare them against the baselines. Not all combinations of session encoders with loss function are presented, only the most promising or interesting ones from the future research perspective (e.g., \texttt{NCAS} for \texttt{RSC15/64} and \texttt{RR/5}).

\begin{table}[t!]
    \center
    \scriptsize
    \begin{tabular}{l|lrrr|rr}
Dataset & Method &  MAP &              PREC & $\bigtriangledown$REC &                HR &               MRR \\
\midrule
\multirow{14}{*}{\texttt{SI-T}} & \texttt{SML-TextCNN-Triplet} &               0.0407 &               0.0526 &          \textbf{0.7463} &      \textbf{0.8525} &  \underline{0.6050} \\
   & \texttt{SML-RNN-Triplet} &  \underline{0.0407} &               0.0526 &      \underline{0.7462} &  \underline{0.8523} &               0.6015 \\
   & \texttt{VSKNN} &      \textbf{0.0410} &  \underline{0.0619} &                   0.7455 &               0.8511 &      \textbf{0.6088} \\
   & \texttt{SML-MaxPool-Triplet} &               0.0400 &               0.0518 &                   0.7368 &               0.8414 &               0.5974 \\
   & \texttt{SML-MaxPool-NCAS} &               0.0389 &               0.0504 &                   0.7195 &               0.8246 &               0.5857 \\
   & \texttt{SML-RNN-NCAS} &               0.0389 &               0.0503 &                   0.7193 &               0.8248 &               0.5910 \\
   & \texttt{SML-TextCNN-NCAS} &               0.0386 &               0.0499 &                   0.7152 &               0.8206 &               0.5844 \\
   & \texttt{SPOP} &               0.0336 &               0.0437 &                   0.6384 &               0.7317 &               0.5724 \\
   & \texttt{SML-TagSpace-Triplet} &               0.0335 &               0.0435 &                   0.6352 &               0.7311 &               0.5602 \\
   & \texttt{GRU4Rec+} &               0.0318 &               0.0463 &                   0.5948 &               0.7578 &               0.5437 \\
   & \texttt{SKNN} &               0.0293 &      \textbf{0.4524} &                   0.5609 &               0.6533 &               0.5640 \\
   & \texttt{POP} &               0.0192 &               0.0257 &                   0.3600 &               0.3954 &               0.1369 \\
   & \texttt{MARKOV-1} &               0.0376 &               0.0218 &                   0.2433 &               0.2875 &               0.1965 \\
\midrule
\multirow{13}{*}{\texttt{SI-D}} & \texttt{VSKNN} &      \textbf{0.0394} &  \underline{0.1228} &          \textbf{0.6499} &      \textbf{0.7484} &      \textbf{0.4483} \\
   & \texttt{SML-RNN-Triplet} &  \underline{0.0374} &               0.0536 &      \underline{0.6401} &  \underline{0.7350} &  \underline{0.4468} \\
   & \texttt{SML-TextCNN-Triplet} &               0.0371 &               0.0531 &                   0.6375 &               0.7334 &               0.4445 \\
   & \texttt{SML-MaxPool-NCAS} &               0.0360 &               0.0515 &                   0.6215 &               0.7185 &               0.4379 \\
   & \texttt{SML-RNN-NCAS} &               0.0355 &               0.0509 &                   0.6153 &               0.7130 &               0.4358 \\
   & \texttt{SML-TextCNN-NCAS} &               0.0346 &               0.0496 &                   0.6033 &               0.6990 &               0.4171 \\
   & \texttt{SKNN} &               0.0350 &      \textbf{0.1502} &                   0.5942 &               0.6878 &               0.4321 \\
   & \texttt{SML-MaxPool-Triplet} &               0.0340 &               0.0489 &                   0.5929 &               0.6822 &               0.3718 \\
   & \texttt{SML-TagSpace-Triplet} &               0.0309 &               0.0445 &                   0.5470 &               0.6261 &               0.3733 \\
   & \texttt{SPOP} &               0.0285 &               0.0406 &                   0.5180 &               0.5853 &               0.4263 \\
   & \texttt{GRU4Rec+} &               0.0332 &               0.0680 &                   0.4966 &               0.6450 &               0.3043 \\
   & \texttt{MARKOV-1} &               0.0348 &               0.0515 &                   0.2582 &               0.3038 &               0.1741 \\
   & \texttt{POP} &               0.0122 &               0.0197 &                   0.2040 &               0.2248 &               0.0655 \\
\midrule
\multirow{13}{*}{\texttt{RSC15/64}} & \texttt{SML-RNN-NCAS} &               0.0358 &               0.0639 &          \textbf{0.5248} &  \underline{0.6557} &      \textbf{0.2884} \\
   & \texttt{SML-MaxPool-NCAS} &               0.0355 &               0.0634 &      \underline{0.5213} &               0.6502 &               0.2841 \\
   & \texttt{SML-TextCNN-NCAS} &               0.0351 &               0.0627 &                   0.5145 &               0.6393 &               0.2766 \\
   & \texttt{SML-RNN-Triplet} &               0.0348 &               0.0623 &                   0.5126 &               0.6371 &               0.2824 \\
   & \texttt{VSKNN} &      \textbf{0.0386} &      \textbf{0.0928} &                   0.5009 &      \textbf{0.6961} &  \underline{0.2879} \\
   & \texttt{SML-MaxPool-Triplet} &               0.0337 &               0.0607 &                   0.4975 &               0.6143 &               0.2680 \\
   & \texttt{SKNN} &  \underline{0.0363} &  \underline{0.0881} &                   0.4780 &               0.6423 &               0.2522 \\
   & \texttt{GRU4Rec+} &               0.0285 &               0.0721 &                   0.4009 &               0.6528 &               0.2752 \\
   & \texttt{SML-TextCNN-Triplet} &               0.0244 &               0.0462 &                   0.3793 &               0.4615 &               0.1389 \\
   & \texttt{SML-TagSpace-Triplet} &               0.0218 &               0.0416 &                   0.3398 &               0.4082 &               0.1396 \\
   & \texttt{MARKOV-1} &               0.0333 &               0.0446 &                   0.3011 &               0.3912 &               0.1771 \\
   & \texttt{SPOP} &               0.0164 &               0.0318 &                   0.2879 &               0.3464 &               0.2205 \\
   & \texttt{POP} &               0.0063 &               0.0129 &                   0.1075 &               0.1264 &               0.0292 \\
\midrule
\multirow{13}{*}{\texttt{RR/5}} & \texttt{SKNN} &      \textbf{0.0283} &      \textbf{0.0532} &          \textbf{0.4704} &      \textbf{0.5788} &               0.3370 \\
   & \texttt{SML-MaxPool-NCAS} &               0.0273 &               0.0443 &      \underline{0.4673} &               0.5692 &               0.3340 \\
   & \texttt{VSKNN} &  \underline{0.0278} &  \underline{0.0531} &                   0.4632 &  \underline{0.5745} &               0.3395 \\
   & \texttt{SML-RNN-NCAS} &               0.0270 &               0.0437 &                   0.4609 &               0.5600 &               0.3379 \\
   & \texttt{GRU4Rec+} &               0.0272 &               0.0502 &                   0.4559 &               0.5669 &               0.3237 \\
   & \texttt{SML-RNN-Triplet} &               0.0270 &               0.0456 &                   0.4542 &               0.5560 &  \underline{0.3605} \\
   & \texttt{SML-TextCNN-NCAS} &               0.0264 &               0.0427 &                   0.4516 &               0.5476 &               0.3197 \\
   & \texttt{SML-MaxPool-Triplet} &               0.0245 &               0.0392 &                   0.4357 &               0.5243 &               0.3451 \\
   & \texttt{SPOP} &               0.0201 &               0.0331 &                   0.3773 &               0.4614 &      \textbf{0.3985} \\
   & \texttt{SML-TextCNN-Triplet} &               0.0183 &               0.0293 &                   0.3278 &               0.3988 &               0.2156 \\
   & \texttt{SML-TagSpace-Triplet} &               0.0117 &               0.0192 &                   0.2321 &               0.2817 &               0.2397 \\
   & \texttt{MARKOV-1} &               0.0183 &               0.0428 &                   0.1557 &               0.1964 &               0.1162 \\
   & \texttt{POP} &               0.0009 &               0.0020 &                   0.0181 &               0.0227 &               0.0058 \\

\end{tabular}
    \normalsize
     \caption{Results obtained during the experiments. The baseline \texttt{SKNN}, \texttt{VSKNN} and \texttt{GRU4Rec+} values for \texttt{RR/5} and \texttt{RSC15/64} are taken from supplementary materials to \citet{Ludewig2019}. Best results for each measure--dataset pair are in \textbf{boldface}, while the second bests are \underline{underlined}; $\bigtriangledown$ indicates the sort column. The \texttt{SML} naming convention is: \texttt{SML-SessionEncoder-LossFunction}, with: \texttt{RNN} and \texttt{MaxPool} denote encoders described in \ref{sec:encoders}, and three loss functions: \texttt{Contrastive}, \texttt{Triplet} and smoothed \texttt{NCA} --- \texttt{NCAS}.}
    \label{tab:experimental_results}
\end{table}

The modification introduced by \texttt{VSKNN} to the non-weighted version of the method (i.e., \texttt{SKNN}) seems to be effective for all the datasets, thus making \texttt{VSKNN} a strong baseline indeed. Nevertheless, in some cases (like \texttt{RR/5}), the simpler \texttt{SKNN} method still obtains better results. Dataset specifics and used metrics play an important role here, as can be seen in Fig. \ref{fig:preprocessed_session_lengths} (right)---\texttt{RR/5} in comparison with other datasets (especially \texttt{RSC15/64}) contains more repeating items. If we place them at the beginning of our recommendation and fill up the rest with the most popular items, we can receive high MRR@20 values. However, the practical usefulness of such recommendations can be questionable.

The low results of \texttt{MARKOV-1} for all datasets show that a simple association of the item and the next following action is not enough to obtain good results. Extracting additional information from entire sequences is needed to improve recommendations, which is the basis on which the sequential modeling with \texttt{GRU4Rec+} method stands. Still, in most cases, it is less accurate in the meaning of used metrics than the simple heuristic of \texttt{VSKNN}. One possible explanation is that the \texttt{VSKNN} model additionally incorporates \emph{recency} in the scoring function. We can consider that as a simply encoded contextual information about when the sequence occurred. This information is not used in other models. When scoring sequences within short periods of time this may not introduce a big difference, but becomes important as the time difference increases, as e.g., some trends arise, and others fade out.

From the overall results, our SML family methods are the best for two datasets, the proprietary \texttt{SI-T} and the open available \texttt{RSC15/64}. For \texttt{SI-T} the proposed triplet loss function seems to be the right choice, wherein the case of \texttt{RSC15/64}, training with \texttt{NCAS} is more stable and is giving overall better results. This situation can be caused by far bigger inventory size and number of events in this dataset. Moreover, on \texttt{SI-D} and \texttt{RR/5} our methods position themselves as the second-best ones with a minimal margin to kNN based methods, \texttt{VSKNN} and \texttt{SKNN}, respectively. For \texttt{SI-D} only the PREC@20 is lower, due to the fact of far better results of \texttt{SKNN} (which we double-checked for the correctness with such good results for both SI datasets). The Retail-Rocket dataset presents consistent results with \citet{Ludewig2019}, where many new methods cope to beat \texttt{SKNN}. With \texttt{SML-MaxPooling-NCAS}, we get close to the position of being the leader.

Between the investigated encoders, we can observe from the results that a simple max-pooling performs well and falls very close to the best score for \texttt{SI-*} datasets. Intuitively, GRU and CNN based methods should be better in encoding longer sequences of actions, like \texttt{RSC15/64} and \texttt{RR/5} (see Fig. \ref{fig:preprocessed_session_lengths} (left)). However, this proved to be true only for \texttt{RSC15/64} results, where CNN and RNN based methods are among the best ones. For \texttt{RR/5} simple pooling with the proposed \texttt{NCAS} loss function is the best one from the SML method family. Additionally, in practical terms, CNN-based models can be preferred from GPU utilization perspective, as the architecture and many libraries are optimized for computer vision and image processing.

\subsection{Coverage and popularity bias} Similar to \cite{Ludewig2018eval,Ludewig2019} we investigated the distribution of predicted items for the selected approaches. Interestingly, our metric learning based methods usually give wider spectrum of recommended items. Even checking simple statistic of overall unique items being recommended, for \texttt{SI} datasets our methods return almost twice as much unique items as \texttt{VSKNN} method (666 to 1,542 and 1,522 to 2,542 for a sample run, all items 2k, 3k respectively, see Table~\ref{tab:datasets}), while for \texttt{RR/5} and \texttt{RSC} the difference is not so big (16,334 to 19,063, 12,232 to 11,216 for a sample run).

\subsection{Ablation Study} To verify the impact of each component in our proposed solution, we run a series of experiments on \texttt{SI-T} dataset for two encoders: \texttt{RNN} and \texttt{MaxPool}, enabling each improvements one by one. The results with REC@20 and MRR@20 are shown in Table~\ref{tab:ablation_results}. 

\begin{table}[t]
    \centering
    \small
    \begin{tabular}{ccl|rr|rr}
Comm. &&& \multicolumn{2}{c|}{\texttt{RNN}} & \multicolumn{2}{c}{\texttt{MaxPool}}\\
Emb. & Sampler &      Loss & $\bigtriangledown$\texttt{REC@20} &  \texttt{MRR@20} & \texttt{REC@20} &  \texttt{MRR@20} \\
\midrule
 True & Pos--Neg  & N-M    & 0.7435 &  0.5973 &  0.7402 &  0.5978 \\
 True & Pos--Neg  & N      & 0.7377 &  0.5973 &  0.7340 &  0.5932 \\
 True & Pos--Neg  & N-M-S  & 0.7371 &  0.5908 &  0.7359 &  0.5888 \\
False & Pos--Neg & N       & 0.6565 &  0.5746 &  0.6508 &  0.5727 \\
False & Pos--Neg & N-M     & 0.6341 &  0.5783 &  0.6192 &  0.5767 \\
False & Pos--Neg & N-M-S   & 0.6192 &  0.5678 &  0.6247 &  0.5796 \\
False & SW       & N-M-S   & 0.0022 &  0.0006 &  0.0525 &  0.0191 \\
\end{tabular}


    \vspace{0.5em}
    \normalsize
    \caption{Ablation results obtained for the \texttt{RNN} and \texttt{MaxPool} session encoders. Columns labels in order: (1) \texttt{True/False} is common embedding was used; (2) Sampler: \texttt{SW} - sliding window, \texttt{Pos-Neg} - session positive negative sampling as described in sec.~\ref{sec:sampling}; (3) Triplet loss with: \texttt{N} - L2 normalization, \texttt{M} - 0.3 margin used, \texttt{S} - swaping anchor--session with positive item. Resultes are sorted by REC@20. }
    \label{tab:ablation_results}
\end{table}

One of the first sampling methods evaluated with \texttt{SML} was a simple sliding window-based technique. For a defined number of events (padded if necessary), we take only the next following items as positive examples, and negative ones are randomly sampled. We quickly switched to sampling presented in section~\ref{sec:sampling}, as we notice that the windowing technique is not reflecting how the system is utilized in real use cases. Specifically, for various sub-sequences from the beginning of a session, predictions are also required, disregarding the sliding window size. As the next step, we evaluated the impact of the inner elements from triplet loss, like normalization (which is very common), margin usage (which for some datasets are set to very small values), and swapping of anchor and positive elements. To our surprise, swapping is not always giving good results for a session-based recommendations setting. 

A crucial role for improving our model was the use of common embeddings for both session encoder $\varphi(s_k)$ and items encoder $\omega(i_j)$ for the prediction. This lowered the number or all parameters to train and positively influenced the overall results. We think that even further improvements can be made to the proposed method by a more global network parameters search. But this was out of scope of our computational possibilities. Thus, we constrained some of the network's hyper-parameters that are related (e.g., GRU hidden state dimension and following feed-forward network dimension to be the same).

\section{Conclusions}

In this paper, we have presented a novel approach to session-based recommendations that utilizes concepts from the field of metric learning. The proposed method has a clear and modular architecture that combines session and item embeddings with a metric loss function. Each of these elements may be individually tweaked and thus defines a potential direction for further research. We test our approach against independent results obtained for strong baseline methods using a well-established evaluation procedure and receive state-of-the-art results. The analysis is also extended by ablation studies, which confirm that the proposed solution does not have unnecessary elements.

Our approach's main advantage is a modular design and extensibility that makes it possible to tweak its components to best match the dataset or incorporate some prior knowledge. Moreover, the fact that \texttt{SML} is based on principles originating from metric learning, many improvements from that field can still be transferred and evaluated for session-based recommendations. From usage perspective, our approach can be attractive in combination with existing pipelines (KNN recommendations) and libraries (optimized CNN).

We can identify two main weaknesses of our method.  Firstly, sampling has a significant impact on the results both in terms of quality and computational efficiency, so careful GPU usage and memory management is required. Secondly, many improvements that can be taken for granted within computer vision do not necessarily improve the final model when combined with other elements for session-based recommendations, which was presented in the ablation study.

Although we achieved promising results with the current method, this work only touched the subject of applying metric learning to session-based recommendations, and much more is to be explored. Apart from the already mentioned embeddings, the positive/negative sampling strategy used during the training phase seems to deserve more attention. Based on good experimental results achieved by some baselines, an introduction of a missing users' actions time context into session-based recommendation also seems worth exploring. Further investigation of improvements in the deep metric learning field can result in even better session-based recommendations, and similar synergy can be found, like in the case of NLP.

\section*{Acknowledgments and Disclosure of Funding}
We acknowledge the support from Sales Intelligence and co-funding by European Regional Development Fund, project number: POIR.01.01.01-00-0632/18. 

\bibliographystyle{unsrt}  
\bibliography{ref}

\begin{thebibliography}{10}

\bibitem{Hidasi}
Bal{\'{a}}zs Hidasi, Alexandros Karatzoglou, Linas Baltrunas, and Domonkos
  Tikk.
\newblock {Session-Based Recommendations with Recurrent Neural Networks}.
\newblock Technical report.

\bibitem{twardowski2016modelling}
Bart{\l}omiej Twardowski.
\newblock Modelling contextual information in session-aware recommender systems
  with neural networks.
\newblock In {\em Proceedings of the 10th ACM Conference on Recommender
  Systems}, pages 273--276, 2016.

\bibitem{Smirnova2017rnn}
Elena Smirnova and Flavian Vasile.
\newblock Contextual sequence modeling for recommendation with recurrent neural
  networks.
\newblock In {\em Proceedings of the 2nd Workshop on Deep Learning for
  Recommender Systems}, DLRS 2017, page 2–9, New York, NY, USA, 2017.
  Association for Computing Machinery.

\bibitem{devlin2018bert}
Jacob Devlin, Ming-Wei Chang, Kenton Lee, and Kristina Toutanova.
\newblock Bert: Pre-training of deep bidirectional transformers for language
  understanding.
\newblock {\em arXiv preprint arXiv:1810.04805}, 2018.

\bibitem{sun2019bert4rec}
Fei Sun, Jun Liu, Jian Wu, Changhua Pei, Xiao Lin, Wenwu Ou, and Peng Jiang.
\newblock Bert4rec: Sequential recommendation with bidirectional encoder
  representations from transformer.
\newblock In {\em Proceedings of the 28th ACM International Conference on
  Information and Knowledge Management}, pages 1441--1450, 2019.

\bibitem{Ludewig2019}
Malte Ludewig, Noemi Mauro, Sara Latifi, and Dietmar Jannach.
\newblock {Performance comparison of neural and non-neural approaches to
  session-based recommendation}.
\newblock In {\em RecSys 2019 - 13th ACM Conference on Recommender Systems},
  pages 462--466. Association for Computing Machinery, Inc, sep 2019.

\bibitem{hidasi2016general}
Bal{\'a}zs Hidasi and Domonkos Tikk.
\newblock General factorization framework for context-aware recommendations.
\newblock {\em Data Mining and Knowledge Discovery}, 30(2):342--371, 2016.

\bibitem{Rendle2010FM}
Steffen Rendle.
\newblock {Factorization machines}.
\newblock In {\em Proceedings - IEEE International Conference on Data Mining,
  ICDM}, pages 995--1000, 2010.

\bibitem{Koren2009td}
Yehuda Koren.
\newblock Collaborative filtering with temporal dynamics.
\newblock In {\em Proceedings of the 15th ACM SIGKDD international conference
  on Knowledge discovery and data mining}, pages 447--456, 2009.

\bibitem{Rendle2010FMC}
Steffen Rendle, Christoph Freudenthaler, and Lars Schmidt-Thieme.
\newblock Factorizing personalized markov chains for next-basket
  recommendation.
\newblock In {\em Proceedings of the 19th International Conference on World
  Wide Web}, WWW ’10, page 811–820, New York, NY, USA, 2010. Association
  for Computing Machinery.

\bibitem{Jannach2017}
Dietmar Jannach and Malte Ludewig.
\newblock When recurrent neural networks meet the neighborhood for
  session-based recommendation.
\newblock In {\em Proceedings of the Eleventh ACM Conference on Recommender
  Systems}, RecSys ’17, page 306–310, New York, NY, USA, 2017. Association
  for Computing Machinery.

\bibitem{hidasi2018recurrent}
Bal{\'a}zs Hidasi and Alexandros Karatzoglou.
\newblock Recurrent neural networks with top-k gains for session-based
  recommendations.
\newblock In {\em Proceedings of the 27th ACM International Conference on
  Information and Knowledge Management}, pages 843--852, 2018.

\bibitem{kim2014convolutional}
Yoon Kim.
\newblock Convolutional neural networks for sentence classification.
\newblock {\em arXiv preprint arXiv:1408.5882}, 2014.

\bibitem{Tang2018caser}
Jiaxi Tang and Ke~Wang.
\newblock {Personalized Top-N Sequential Recommendation via Convolutional
  Sequence Embedding}.
\newblock {\em Proceedings of the Eleventh ACM International Conference on Web
  Search and Data Mining - WSDM '18}, pages 565--573, 2018.

\bibitem{Yuan2019}
Fajie Yuan, Alexandros Karatzoglou, Ioannis Arapakis, Joemon~M. Jose, and
  Xiangnan He.
\newblock {A simple convolutional generative network for next item
  recommendation}.
\newblock {\em WSDM 2019 - Proceedings of the 12th ACM International Conference
  on Web Search and Data Mining}, (August):582--590, 2019.

\bibitem{Oord2016}
Aaron van~den Oord, Sander Dieleman, Heiga Zen, Karen Simonyan, Oriol Vinyals,
  Alex Graves, Nal Kalchbrenner, Andrew Senior, and Koray Kavukcuoglu.
\newblock {WaveNet: A Generative Model for Raw Audio}.
\newblock sep 2016.

\bibitem{Ludewig2018eval}
Malte Ludewig and Dietmar Jannach.
\newblock Evaluation of session-based recommendation algorithms.
\newblock {\em User Modeling and User-Adapted Interaction}, page 331–390,
  December 2018.

\bibitem{schultz2004learning}
Matthew Schultz and Thorsten Joachims.
\newblock Learning a distance metric from relative comparisons.
\newblock In {\em Advances in neural information processing systems}, pages
  41--48, 2004.

\bibitem{mcfee2010metric}
Brian McFee and Gert~R Lanckriet.
\newblock Metric learning to rank.
\newblock In {\em Proceedings of the 27th International Conference on Machine
  Learning (ICML-10)}, pages 775--782, 2010.

\bibitem{lim2014efficient}
Daryl Lim and Gert Lanckriet.
\newblock Efficient learning of mahalanobis metrics for ranking.
\newblock In {\em International conference on machine learning}, pages
  1980--1988, 2014.

\bibitem{hadsell2006dimensionality}
Raia Hadsell, Sumit Chopra, and Yann LeCun.
\newblock Dimensionality reduction by learning an invariant mapping.
\newblock In {\em Proceedings of the IEEE Conference on Computer Vision and
  Pattern Recognition (CVPR)}, pages 1735--1742, 2006.

\bibitem{Hoffer2015}
Elad Hoffer and Nir Ailon.
\newblock {Deep metric learning using triplet network}.
\newblock {\em Lecture Notes in Computer Science (including subseries Lecture
  Notes in Artificial Intelligence and Lecture Notes in Bioinformatics)}, pages
  84--92, 2015.

\bibitem{wang2017angular}
Jian Wang, Feng Zhou, Shilei Wen, Xiao Liu, and Yuanqing Lin.
\newblock Deep metric learning with angular loss.
\newblock In {\em Proceedings of the IEEE International Conference on Computer
  Vision}, pages 2593--2601, 2017.

\bibitem{yuan2019signal}
Tongtong Yuan, Weihong Deng, Jian Tang, Yinan Tang, and Binghui Chen.
\newblock Signal-to-noise ratio: A robust distance metric for deep metric
  learning.
\newblock In {\em Proceedings of the IEEE Conference on Computer Vision and
  Pattern Recognition}, pages 4815--4824, 2019.

\bibitem{wang2019multi}
Xun Wang, Xintong Han, Weilin Huang, Dengke Dong, and Matthew~R Scott.
\newblock Multi-similarity loss with general pair weighting for deep metric
  learning.
\newblock In {\em Proceedings of the IEEE Conference on Computer Vision and
  Pattern Recognition}, pages 5022--5030, 2019.

\bibitem{hermans2017defense}
Alexander Hermans, Lucas Beyer, and Bastian Leibe.
\newblock In defense of the triplet loss for person re-identification.
\newblock {\em arXiv preprint arXiv:1703.07737}, 2017.

\bibitem{sohn2016npairs}
Kihyuk Sohn.
\newblock Improved deep metric learning with multi-class n-pair loss objective.
\newblock In {\em Advances in Neural Information Processing Systems}, pages
  1857--1865, 2016.

\bibitem{oh2016lifed}
Hyun Oh~Song, Yu~Xiang, Stefanie Jegelka, and Silvio Savarese.
\newblock Deep metric learning via lifted structured feature embedding.
\newblock In {\em Proceedings of the IEEE Conference on Computer Vision and
  Pattern Recognition}, pages 4004--4012, 2016.

\bibitem{wu2017sampling}
Chao-Yuan Wu, R~Manmatha, Alexander~J Smola, and Philipp Krahenbuhl.
\newblock Sampling matters in deep embedding learning.
\newblock In {\em Proceedings of the IEEE International Conference on Computer
  Vision}, pages 2840--2848, 2017.

\bibitem{fehervari2019unbiased}
Istvan Fehervari, Avinash Ravichandran, and Srikar Appalaraju.
\newblock Unbiased evaluation of deep metric learning algorithms.
\newblock {\em arXiv preprint arXiv:1911.12528}, 2019.

\bibitem{musgrave2020realitycheck}
Kevin Musgrave, Serge Belongie, and Ser-Nam Lim.
\newblock A metric learning reality check.
\newblock {\em arXiv preprint arXiv:2003.08505}, 2020.

\bibitem{kaya2019dmlsurvey}
Mahmut Kaya and Hasan~{\c{S}}akir Bilge.
\newblock Deep metric learning: a survey.
\newblock {\em Symmetry}, 11(9):1066, 2019.

\bibitem{Burges2007}
C~J~C Burges, Robert Ragno, and Q~V Le.
\newblock {Learning to Rank with Nonsmooth Cost Functions}.
\newblock {\em Machine Learning}, 19:193--200, 2007.

\bibitem{Yue2007}
Yisong Yue, Thomas Finley, Filip Radlinski, and Thorsten Joachims.
\newblock A support vector method for optimizing average precision.
\newblock In {\em Proceedings of the 30th annual international ACM SIGIR
  conference on Research and development in information retrieval}, pages
  271--278, 2007.

\bibitem{Taylor2008}
Michael Taylor, John Guiver, Stephen Robertson, and Tom Minka.
\newblock {SoftRank: optimizing non-smooth rank metrics}.
\newblock In {\em WSDM '08}, pages 77--86, 2008.

\bibitem{Ning2011}
Xia Ning and George Karypis.
\newblock {SLIM: Sparse LInear Methods for top-N recommender systems}.
\newblock {\em Proceedings - IEEE International Conference on Data Mining,
  ICDM}, pages 497--506, 2011.

\bibitem{Koren2008}
Yehuda Koren.
\newblock {Factorization meets the neighborhood: a multifaceted collaborative
  filtering model}.
\newblock {\em Proceeding of the 14th ACM SIGKDD international conference on
  Knowledge discovery and data mining}, pages 426--434, 2008.

\bibitem{Rendle2009}
Steffen Rendle, Christoph Freudenthaler, Zeno Gantner, and Lars Schmidt-thieme.
\newblock {BPR : Bayesian Personalized Ranking from Implicit Feedback}.
\newblock In {\em Proceedings of the Twenty-Fifth Conference on Uncertainty in
  Artificial Intelligence}, volume cs.LG, pages 452--461, 2009.

\bibitem{Liu2009}
Nn~Liu, Min Zhao, and Q~Yang.
\newblock {Probabilistic latent preference analysis for collaborative
  filtering}.
\newblock {\em Proceedings of the 18th ACM conference on on Information and
  knowledge management}, pages 759--766, 2009.

\bibitem{Chen2006}
Harr Chen and David~R. Karger.
\newblock {Less is More: Probabilistic Models for Retrieving Fewer Relevant
  Documents}.
\newblock In {\em SIGIR}, pages 429 -- 436, 2006.

\bibitem{Shi2013}
Yue Shi, Alexandros Karatzoglou, and Linas Baltrunas.
\newblock {xCLiMF: optimizing expected reciprocal rank for data with multiple
  levels of relevance}.
\newblock {\em Proceedings of the 7th ACM conference on Recommender systems},
  pages 0--3, 2013.

\bibitem{GAPfm}
Yue Shi, Alexandros Karatzoglou, Linas Baltrunas, Martha Larson, and Alan
  Hanjalic.
\newblock {GAPfm}.
\newblock In {\em Proceedings of the 22nd ACM international conference on
  Conference on information {\&} knowledge management - CIKM '13}, pages
  2261--2266, 2013.

\bibitem{Shi2012}
Yue Shi, Alexandros Karatzoglou, Linas Baltrunas, Martha Larson, Alan Hanjalic,
  and Nuria Oliver.
\newblock {TFMAP: Optimizing MAP for top-n context-aware recommendation}.
\newblock {\em Proceedings of the 35th international ACM SIGIR conference on
  Research and development in information retrieval}, pages 155--164, 2012.

\bibitem{liu2009learning}
Tie-Yan Liu et~al.
\newblock Learning to rank for information retrieval.
\newblock {\em Foundations and Trends in Information Retrieval}, 2009.

\bibitem{weston2011wsabie}
Jason Weston, Samy Bengio, and Nicolas Usunier.
\newblock Wsabie: Scaling up to large vocabulary image annotation.
\newblock In {\em Twenty-Second International Joint Conference on Artificial
  Intelligence}, 2011.

\bibitem{goldberger2005nca}
Jacob Goldberger, Geoffrey~E Hinton, Sam~T Roweis, and Russ~R Salakhutdinov.
\newblock Neighbourhood components analysis.
\newblock In {\em Advances in neural information processing systems}, pages
  513--520, 2005.

\bibitem{movshovitz2017proxy-nofuss}
Yair Movshovitz-Attias, Alexander Toshev, Thomas~K Leung, Sergey Ioffe, and
  Saurabh Singh.
\newblock No fuss distance metric learning using proxies.
\newblock In {\em Proceedings of the IEEE International Conference on Computer
  Vision}, pages 360--368, 2017.

\bibitem{weston-etal-2014-tagspace}
Jason Weston, Sumit Chopra, and Keith Adams.
\newblock {\#}{T}ag{S}pace: Semantic embeddings from hashtags.
\newblock In {\em Proceedings of the 2014 Conference on Empirical Methods in
  Natural Language Processing ({EMNLP})}, pages 1822--1827, Doha, Qatar,
  October 2014. Association for Computational Linguistics.

\bibitem{pytorch}
Adam Paszke, Sam Gross, Francisco Massa, Adam Lerer, James Bradbury, Gregory
  Chanan, Trevor Killeen, Zeming Lin, Natalia Gimelshein, Luca Antiga, Alban
  Desmaison, Andreas Kopf, Edward Yang, Zachary DeVito, Martin Raison, Alykhan
  Tejani, Sasank Chilamkurthy, Benoit Steiner, Lu~Fang, Junjie Bai, and Soumith
  Chintala.
\newblock Pytorch: An imperative style, high-performance deep learning library.
\newblock In H.~Wallach, H.~Larochelle, A.~Beygelzimer, F.~d~Alch\'{e}-Buc,
  E.~Fox, and R.~Garnett, editors, {\em Advances in Neural Information
  Processing Systems 32}, pages 8024--8035. Curran Associates, Inc., 2019.

\end{thebibliography}

\end{document}